\documentclass[%
 reprint,
 amsmath,amssymb,
 aps,
]{revtex4-2}

\usepackage{graphicx}
\usepackage{dcolumn}
\usepackage{bm}
\usepackage{soul,framed} 
\usepackage{subfigure}
\usepackage{fancyhdr}
\usepackage{booktabs}
\usepackage{tabularx}
\usepackage{comment}

\newcommand{\Fig}[1]{Fig.\,{\ref{#1}}}
\newcommand{\Figs}[1]{Figs.~$\!${\ref{#1}}}

\usepackage{amsmath,amssymb,amsfonts}
\usepackage{textcomp}
\usepackage{gensymb}
\usepackage{color}
\usepackage{array}
\usepackage[table,xcdraw]{xcolor}
\usepackage{multirow}

\def\BibTeX{{\rm B\kern-.05em{\sc i\kern-.025em b}\kern-.08em
    T\kern-.1667em\lower.7ex\hbox{E}\kern-.125emX}}

\begin{document}

\title{Space-time Metallic Metasurfaces for Frequency Conversion and Beamforming}

\author{Salvador Moreno-Rodríguez$^{1,*}$}
\affiliation{
$^1$Department of Signal Theory, Telematics and Communications, Research Centre for Information and Communication Technologies (CITIC-UGR), Universidad de Granada, 18071 Granada, Spain}%
\email{Corresponding author: salvamr96@ugr.es}
\author{Antonio Alex-Amor,$^2$ Pablo Padilla,$^1$ Juan F. Valenzuela-Valdés$^1$ and Carlos Molero$^1$}
\affiliation{%
$^2$University of Pennsylvania, Department of Electrical and Systems Engineering, Philadelphia, Pennsylvania 19104, United States
}

\begin{abstract}
    This paper details a class of metal-based space-time metasurfaces for application in wireless communications scenarios. Concretely, we describe space-time metasurfaces that periodically alternate their properties in time between three spatial states: "air", "conductor" and "grating". We analyze the physics of these metastructures via a computationally-efficient analytical technique based on the use of Floquet-Bloch series, integral equations and circuit models. By doing so, we reveal important features of these spatiotemporal metasurfaces: scattering parameters, field profiles, diffraction angles and nature of the space-time harmonics. The results, corroborated with a self-implemented numerical FDTD approach, show the potential application of these space-time metasurfaces as beamformers acting in reflection, in transmission or both. The amplitude and direction of the diffracted orders can be electronically controlled with the paramaters of the metasurface. Moreover, the intrinsic ability of time-modulated diffractive metasurfaces to mix and multiply frequencies is tested. We show how two different modulations can lead to the same diffraction angle but with different mixed output frequencies.
\end{abstract}

\maketitle

\section{Introduction}\label{sec:introduction}

Metamaterials have been conceived as artificially engineered devices with the property of manipulating electromagnetic waves. The term metamaterial originally referred to structures exhibiting negative permittivity and permeability simultaneously \cite{falcone2004effective}. However, in the subsequent years this conception has experimented different evolutions, remarking the emergence of thinner/planar versions called metasurfaces \cite{quevedo2019roadmap}. 

A conventional metasurface can be physically described as a thin screen with planar geometry exhibiting spatial modulation \cite{martini2022theory}. Thanks to this modulation, metasurfaces become excellent terminals for a wide range of applications, such as microwave absorbers \cite{ra2015thin}, beam shaping \cite{pfeiffer2013metamaterial}, or lenses \cite{caloz2004wave}, among others. Currently, researchers are paying close attention to a new evolution of metasurfaces, the so-called \emph{space-time metasurfaces}, in which time is employed as an additional modulation to the existing purely spatial one 
\cite{morgenthaler1958velocity, zurita2009reflection}. This introduces additional degrees of freedom as well as richer physics on the system \cite{caloz2019spacetime, caloz2019spacetimeII, pacheco2022time}. Likewise, time modulation also enhances the capabilities of the active terminals \cite{shaltout2019spatiotemporal}, leading to promising concepts such as intelligent programmable metasurfaces \cite{cui2014coding} that could be combined and integrated into wireless communications and radars systems \cite{pacheco2020temporal}.

The recent development and fabrication of magnetic-free non-reciprocal devices \cite{reiskarimian2016magnetic,taravati2017nonreciprocal,estep2014magnetic,taravati2020full,wu2020space} has been a key factor for the  interest in space-time modulated devices \cite{engheta2023four,mikheeva2022space}. Thus, the number of theoretical studies of space-time varying structures in the microwave and optics range \cite{taravati2019generalized,caloz2019spacetime,galiffi2022photonics} have increased in the last few years favoring the emergence of novel theoretical applications, such as the use of temporal photonic crystals to achieve  amplification \cite{gaxiola2023growing}, filtering and isolation \cite{wu2019isolating}, the utilization of travelling-wave modulations for  power combining \cite{wang2021space}, the use of grounded slab for efficient phase conjugations \cite{yin2022efficient}, production of temporal chirping and lensing \cite{pacheco2023temporal},  or frequency mixing and multiplying \cite{li2023space}. 

Specifically, in the field of telecommunications, radar and wireless systems, frequency mixing and multiplying is of capital relevance \cite{zhou1994efficient,tang2020photonics}. Traditionally, frequency mixing has been achieved via nonlinear circuit components such as Schottky diodes, GaAs FETs and CMOS transistors. Interestingly,  frequency mixing is an intrinsic property of time-modulated diffractive systems, as the frequency of the diffracted waves is directly related to the frequency of the incident wave \cite{liu2018all, taravati2018aperiodic}. Thus, spatiotemporal metasurfaces  can complement the use of nonlinear components and even replace them in frequency ranges or scenarios where they are no longer functional. Time-modulated antennas have opened new alternatives in this regard as well \cite{taravati2016mixer, zang2019nonreciprocal}.

Furthermore, the versatility to simultaneously modify the momentum and frequency of waves has led space-time-modulated devices to be strong candidates for beamforming or beamsteering \cite{taravati2022microwave,wu2022sideband}. Recently, it has been possible to test the beamsteering capabilities of these devices by modifying the phase constant associated to each unit cell with different temporal sequences \cite{vosoughitabar2023programming}. It probes the potential of these metasurfaces as candidates for the future intelligent communications \cite{pacheco2023temporal}.

Unfortunately, the lack of available commercial solutions to simulate complex behavior in space-time-modulated media hampers the analysis and design of these metadevices. At present, commercial simulator COMSOL Multiphysics is the only direct alternative for simulating spatiotemporal metamaterials. Nonetheless, it naturally presents the limitations inherent to a general-purpose electromagnetic simulator. As an in-house alternative, many of these applications are usually simulated using numerical techniques such as the finite-difference time-domain (FDTD) method \cite{stewart2017finite}. However, the high computational costs and long simulation times that this code entails, have prompted the development of adaptive mesh-based solutions to enhance computational efficiency, such as the Discontinuous Galerkin Time-Domain (DGTD) method \cite{zhao2022space}. Others numerical solutions based on generalized sheet transition conditions (GSTCs) \cite{vahabzadeh2017generalized, tiukuvaara2021floquet}, modal techniques \cite{khorrami2021dynamical} and integral-equation methods \cite{rahmanzadeh2023analysis} show high performance when simulating time-varying metasurfaces. Nonetheless, many times the researchers also search approaches that can provide a deep physical insight of the studied problems. Thus, some alternatives based on analytical and semi-analytical techniques have already been reported \cite{scarborough2020accelerated, huidobro2021homogenization, mostafa2022coherently, pham2022diffraction}. 

In this work, we implement an analytical technique for the analysis of space-time metallic metasurfaces that can commute between three states: ``air", ``conductor" and ``grating". As it will be show, a wise temporal combination of the three mentioned spatial states can produce a rich diffraction spectrum. Transmission and reflection levels can be tuned at convenience depending on the amount of time the metasurface is in these states. The air state favours transmission, while the conductor state favours reflection. The addition of the grating state adds richness and variety to the diffraction spectrum. The present space-time metasurface is of potential interest for related applications such as frequency conversion and beamforming.

Our analytical proposal is based on a Floquet-Bloch series expansion of the electromagnetic fields and integral-equation techniques that account for the space-time boundary conditions. The analytical formulation comes with an equivalent circuit that gives phyisical insight on spatiotemporal diffraction. Interestingly, the associated circuit model comes from first-principle computations; namely, the topology and elements of the circuit are not imposed by the designer, but arise as a consequence of the analytical formulation itself. Analytical circuits for spatially-modulated \cite{rodriguez2012analytical, molero2021cross, alex2023analytical} and temporally-modulated \cite{alex2023diffraction, moreno2023time, alex2023analysis} systems  have been previously reported by the authors.  The \emph{analytical circuit} is a physically-insightful object that will give us information on the nature of the space-time harmonics, coupling between them, preferred diffraction directions and scattering parameters for both transverse magnetic (TM) and transverse electric (TE) polarizations. Moreover, computational times are minimal: a wideband range of frequencies can be simulated in less than a second. 

The paper is organized as follows. Section II presents the space-time metasurface and depicts the analytical framework. We discuss on the physics of the diffraction spectrum, field profiles, nature of Floquet harmonics and scattering parameters via the associated equivalent circuit. Section III focuses on describing the feasibility of the spatiotemporal metasurface for use in wireless applications. Concretely, we focus on its beamforming and frequency-mixing capabilities, both studied via the analytical method and a supportive self-implemented FDTD approach. Finally, general conclusions are drawn in Section IV.

\section{Analytical Framework}
\begin{figure}[!t]
	\centering
    \includegraphics[width= 1\columnwidth]{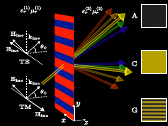}
	\caption{Sketch of the space-time-varying metasurface. The metasurface can alternate in time between three spatial states: air (A), conductor (C) and grating (G). By smartly combining the three states, a rich diffraction phenomenology can be created, from which wireless communications systems can take advantage.}
	\label{fig1}
\end{figure}

\subsection{Space-Time Metasurface}

\Fig{fig1} depicts the spatio-temporal system studied in this manuscript. The system is formed by a space-time-modulated diffractive metasurface that manipulates the incident transverse electric (TE) or transverse magnetic (TM) plane waves. The diffractive metasurface is based on different strips painted in red and blue that alternate their electromagnetic properties in time. Each strip can either become electromagnetically transparent or behave like a solid metal as it is assumed in \cite{stefanini2023temporal}. 

Depending on the electromagnetic behavior of each of the strips, the space-time-modulated diffractive metasurface can operate in \emph{three different states}: ``air'' (A), ``conductor'' (C) and ``grating" (G). In the \emph{air state}, all strips appear to vanish from an electromagnetic perspective, so full transmission of the incident wave is expected. In the \emph{conductor state}, all strips behave as a solid metal achieving a solid metallic wall, so full reflection of the incident wave is expected. In the case of the \emph{grating state}, the red strips behave as solid metals, and the blue strips become electromagnetically transparent, causing the space-time metasurface to operate as a spatially-modulated metallic diffraction grating. 

The three states of the space-time metasurface can be realized by alternating between total transparency or total reflectivity, as recently achieved in \cite{fang2023multifunctional}. Nonetheless, the implementation of the metasurface could also be inspired by several innovative prototypes in the microwave range \cite{ataloglou2023metasurfaces}. Many of these devices are based on adjustable biased PIN diodes \cite{zhang2018space, zhang2019dynamically,fang2023diverse} and varactors \cite{wu2020space,cardin2020surface, taravati2020full}, traditionally implemented in reflectarray and trasmittarray systems. Moreover, novel theoretical designs based on the use of electronically-reconfigurable materials such as graphene \cite{liu2016time,wang2020theory} or transparent conductive oxides \cite{kinsey2015epsilon}   open new alternatives for the millimeter-wave and terahertz regimes.

The geometrical parameters of the metasurface are the following. The space-time diffractive metasurface is placed along the XY plane, as \Fig{fig1} illustrates.  In the time intervals in which the metasurface is in the grating state, it turns into a periodic arrangement of perfect electric conductor (PEC) strips (period $P$ and slit width $W$) that are infinitely-extended along the $x$ axis.  The spatial periodicity of the grating state is along the $y$ axis. In the conductor and air states, the metasurface can be modeled as infinitely-extended uniform thin PEC  and air screens, respectively. The air state assumes electrical parameters $\varepsilon_r = \mu_r = 1$. In order to construct a proper analytical framework, we only need to enforce the temporal variation of the system to follow a \emph{time-periodic} scheme, of period $T_\mathrm{M}$. 

By wisely combining the three states in time (A-G, C-G or A-C), a rich diffraction phenomenology can be created. From an engineering perspective, the diffraction pattern can be electrically tuned by setting the space-time parameters of the system. This enables an efficient beamforming platform that could be potentially applied in wide range of wireless communications scenarios. Remarkably, the space-time metasurface can act as a beamformer in reflection, transmission or both simultaneously by simply tuning its electrical parameters. This will be illustrated in detail in the next sections.
 
\subsection{Floquet-Bloch Expansion}
Since the diffractive metasurface behaves as space-time-periodic media, the transverse electric and magnetic fields in regions (1) and (2) can be expressed in terms of Floquet-Bloch series. Both media are electromagnetically defined by $\varepsilon_{\text{r}}^{(i)}$ and $\mu_{\text{r}}^{(i)}$. Thus, let us consider a space-time metasurface with spatial period $P$ along the $y$-direction and temporal \emph{macroperiod} $T_{\text{M}}$ 
\cite{moreno2023time}. It receives a plane wave impinging obliquely with angular frequency $\omega_{0}$ and TE polarization. The fields at both sides of the metasurface can be expressed in terms of a Floquet expansion of harmonics. In region (1) ($z < 0$):
\begin{multline}\label{ETE1}
    \mathbf{E}_{\text{t}}^{\text{TE},(1)}(y, z, t) =  \bigg[\text{e}^{\text{j}( \omega_{0} t -  k_{0} y - \beta_{00}^{(1)} z)} + R \text{e}^{\text{j} (\omega_{0} t -  k_{0} y + \beta_{00}^{(1)} z)} \\ +  \displaystyle\sum _{\substack{\forall mn \neq 00}} E_{mn}^{(1)} \text{e}^{\text{j} (\omega_{n} t -  k_{m} y + \beta_{mn}^{(1)} z)} \bigg] \hat{\mathbf{x}}
\end{multline}
\begin{multline}\label{HTE1}
    \mathbf{H}_{\text{t}}^{\text{TE},(1)}(y, z, t) =  \bigg[ Y_{00}^{\text{TE},(1)}\text{e}^{\text{j} (\omega_{0} t -  k_{0} y - \beta_{00}^{(1)} z)} \\ - R Y_{00}^{\text{TE},(1)} \text{e}^{\text{j} (\omega_{0} t -  k_{0} y + \beta_{00}^{(1)} z)} \\ - \displaystyle\sum _{\substack{\forall mn \neq 00}} Y_{mn}^{\text{TE},(1)} E_{mn}^{(1)} \text{e}^{\text{j} (\omega_{n} t -  k_{m} y + \beta_{mn}^{(1)} z)} \bigg]\hat{\mathbf{y}}\, .
\end{multline}
In the former expressions, $E_{mn}^{(1)}$ is the amplitude of the $(m,n)$-th space-time harmonic operating in region (1), $R$ is the reflection coefficient associated with the incident wave ($00$-th harmonic), $\omega_{n}$ is the angular frequency associated with the temporal $n$-th harmonic,
\begin{equation} \label{wn}
    \omega_{n} = \omega_{0} + n\frac{2\pi}{T_{\text{M}}} ,
\end{equation}
and $k_{m}$ is the transverse wavenumber linked to the spatial $m$-th harmonic:
\begin{equation} \label{kt_m}
    k_{m} = k_t + m\frac{2\pi}{P}
\end{equation}
with
\begin{equation}    
 k_t = \sqrt{\varepsilon_r^{(1)}\mu_r^{(1)}} \, \frac{\omega_{0}}{c}\sin{(\theta_0)}
\end{equation}
and $\theta_{0}$ being the incidence angle. 
 
In a similar way, we define the tangential electromagnetic field via Floquet expansion in the region $(2)$ ($z > 0$), 
\begin{multline}\label{ETE2}
    \mathbf{E}_{\text{t}}^{\text{TE},(2)}(y, z, t) = \bigg[ T \text{e}^{\text{j} (\omega_{0} t -  k_{0} y - \beta_{00}^{(2)} z)} \\ + \displaystyle\sum _{\substack{\forall mn \neq 00}} E_{mn}^{(2)} \text{e}^{\text{j} (\omega_{n} t -  k_{m} y - \beta_{mn}^{(2)} z)} \bigg] \hat{\mathbf{x}}
\end{multline}
\begin{multline}\label{HTE2}
    \mathbf{H}_{\text{t}}^{\text{TE},(2)}(y, z, t) =  \bigg[ T Y_{00}^{\text{TE},(2)} \text{e}^{\text{j} (\omega_{0} t -  k_{0} y - \beta_{00}^{(2)} z)} \\ +  \displaystyle\sum _{\substack{\forall mn \neq 00}} Y_{mn}^{\text{TE},(2)} E_{mn}^{(2)} \text{e}^{\text{j} (\omega_{n} t -  k_{m} y - \beta_{mn}^{(2)} z)} \bigg]\hat{\mathbf{y}}
\end{multline}
where $T$ denotes the transmission coefficient and $E_{mn}^{(2)}$ the amplitude associated to the $(m,n)$-th  Floquet harmonic.
In the magnetic fields expansions in \eqref{HTE1} and \eqref{HTE2}, $Y_{mn}^{(i)}$ denote the admittance of the the $(m,n)$-th Floquet harmonic located at the $i$-th medium, defined for TE incidence as
\begin{equation}  \label{admittanceTE} 
Y_{mn}^{\text{TE},(i)} = \frac{\beta_{mn}^{(i)}}{\mu_r^{(i)} \mu_0\, \omega_n}, \hspace{5 mm} 
\end{equation}
with 
\begin{align}
    \label{beta} \beta_{mn}^{(i)} &= 
    \sqrt{\varepsilon_r^{(i)}\mu_r^{(i)} \left[\frac{\omega_{n}}{c}\right]^2 - \left[k_{m}\right]^2} \, .
\end{align}
being the propagation constant of the $(m,n)$-th harmonic in region $(i)$.

The previous derivation can be easily adapted for TM incidence. In the TM case, the electric and magnetic field orientations are the opposite to the TE case, thus 
\begin{align}
\mathbf{E}_{\text{t}}^{\text{TM},(i)}(y, z, t) &= [\hat{\mathbf{x}} \cdot \mathbf{E}_{\text{t}}^{\text{TE},(i)}(y, z, t)] \hat{\mathbf{y}} \\
\label{HTM}
\mathbf{H}_{\text{t}}^{\text{TM},(i)}(y, z, t) &= [\hat{\mathbf{y}} \cdot \mathbf{H}_{\text{t}}^{\text{TE},(i)}(y, z, t)] \hat{\mathbf{x}}\,.
\end{align}
In addition, the admittance expressions must now be replaced to those related to TM harmonics,   
\begin{equation}\label{admittanceTM} 
Y_{mn}^{\text{TM},(i)} = \frac{\varepsilon_r^{(i)} \varepsilon_0\, \omega_n}{ \beta_{mn}^{(i)}}\,.
\end{equation}
\subsection{Nature of the Space-Time Harmonics}

 Real(imaginary) values of $\beta_{mn}^{(i)}$ imply that the nature of the $(m,n)$-th Floquet harmonic is propagative(evanescent).  By using eq. \eqref{beta}, it can be shown that all evanescent space-time harmonics fulfill the general conditions
\begin{subequations} \label{m_n}
    \begin{equation}
    \frac{\sqrt{\varepsilon_r \mu_r}}{c} \left(\omega_0   \sin (\theta_0) - \left| \omega_0 + n\frac{2\pi}{T_{\text{M}}} \right| \right) + m \frac{2\pi}{P} > 0 
    \end{equation}
    \begin{equation}
        \frac{\sqrt{\varepsilon_r \mu_r}}{c} \left(\omega_0   \sin (\theta_0) + \left| \omega_0 + n\frac{2\pi}{T_{\text{M}}} \right| \right) + m \frac{2\pi}{P} < 0 
    \end{equation}
\end{subequations}

\noindent In a simpler scenario where the surrounding media are air ($\varepsilon_r = \mu_r = 1$) and the waves impinge normally ($\theta_0 = 0$) to the space-time metasurface, eq. \eqref{m_n} reduces to:

\begin{subequations}
    \begin{equation}\label{Nature_a}
        - \left|1 +  n\frac{T_0}{T_{\text{M}}} \right| +  m \frac{\lambda_0}{P} > 0
    \end{equation}
    \begin{equation}\label{Nature_b}
        +\left|1 +  n\frac{T_0}{T_{\text{M}}}\right| +  m \frac{\lambda_0}{P}  < 0
    \end{equation}
\end{subequations}

\begin{figure}[!t]
	\centering
        \includegraphics[width= 1\columnwidth]{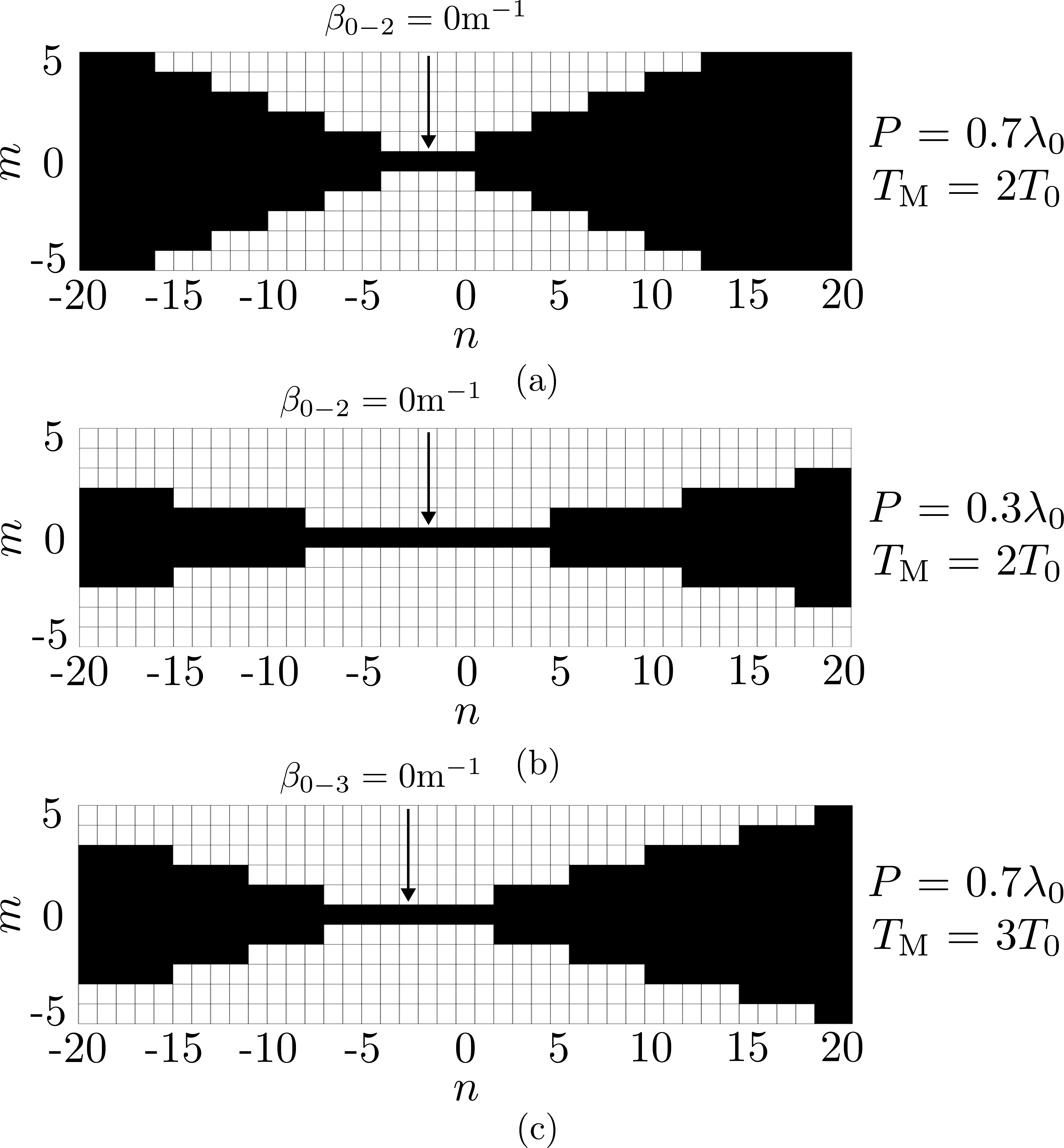}
	\caption{Propagative (black) and evanescent (white) nature of $(m,n)$-harmonics when an incident wave of angular frequency $\omega_0$ = $2\pi \cdot 30$ GHz impinges on the space-time-varying metasurface with: (a) $P$ = $0.7\lambda_0$ and $T_{\text{M}}$ = $2T_0$, (b) $P$ = $0.3\lambda_0$ and $T_{\text{M}}$ = $2T_0$, (c) $P$ = $0.7\lambda_0$ and $T_{\text{M}}$ = $3T_0$. Normal incidence and  $\varepsilon_r = \mu_r = 1$ are assumed. }
	\label{Evanescentes}
\end{figure}

\noindent The former expressions show that, for a fixed temporal macroperiod $T_{\text{M}}$,  there will be a greater number of evanescent space-time harmonics as the spatial periodicity $P$ reduces. Likewise, for a fixed $P$,  there will be a greater number of evanescent harmonics as $T_{\text{M}}$ increases. Moreover, all the $(0,n)$-th harmonics ($m=0, \forall n$) are propagative under normal incidence. This is consistent with the results extracted in our previous work \cite{alex2023diffraction}, as the  $(0,n)$-th Floquet harmonics are only driven by the time modulation.

\Fig{Evanescentes} summarizes all the aforementioned phenomenology related to the space-time harmonics. In the figure, the vertical axis represents the spatial ($m$-indexed) harmonics and the horizontal axis represents the temporal ($n$-indexed) harmonics. The evanescent harmonics are marked in white and the propagative ones in black. Upon analyzing \Fig{Evanescentes}, it is evident the existence of certain symmetry for both indexes. In the case of considering normal incidence, this symmetry appears around the null of the propagation constant in both the vertical and horizontal planes. Therefore, seeking the roots of eq. \eqref{beta} under normal incidence, the $(m,n)$-th harmonic with $\beta_{mn}=0$  emerges for $m=0$ and  $n=-\frac{T_{\text{M}}}{T_0}$. It corresponds to a DC harmonic ($\omega_n= 0$) \cite{tiukuvaara2021floquet}, which is present at ($m=0,n=-2$) in \Fig{Evanescentes}(a)-(b) and at ($m=0,n=-3$) in \Fig{Evanescentes}(c). 

The phenomenology of the propagative and evanescent harmonics is much richer here than in the case considered in our previous works \cite{alex2023diffraction, moreno2023time}. In this work, we are considering a space-time modulation, where in \cite{alex2023diffraction, moreno2023time} we considered a time-only modulation. Time-only modulations led to scenarios where most of the harmonics are propagative, and only a few are evanescent. In fact, it was shown in \cite{alex2023diffraction} that all time harmonics are propagative under normal incidence conditions in the time-modulated metasurface. The situation is rather different and more interesting from a physics and engineering perspective when spatial and temporal modulations intervene together \cite{alex2023analysis}. 

\subsection{Diffraction Angles and Beamforming} \label{Angulos}
The diffraction angle of the $(m,n)$-th Floquet harmonic can be computed using:

\begin{subequations}
    \begin{equation}\label{angle_a}
    \theta_{mn}^{(i)} = \mathrm{asin}\left(\dfrac{k_{m}}{\sqrt{\varepsilon_r^{(i)}\mu_r^{(i)}} \left[\frac{\omega_{n}}{c}\right]}\right) = \mathrm{atan}\left( \frac{k_m}{\beta_{mn}} \right)
    \end{equation}
    \begin{equation}\label{angle_b}
    \theta_{mn}^{(i)} = \mathrm{asin}\left(\dfrac{\sqrt{\varepsilon_r^{(1)}\mu_r^{(1)}} \text{sin}(\theta_0)+m\frac{\lambda_0}{P}}{\sqrt{\varepsilon_r^{(i)}\mu_r^{(i)}}\left[1+n\frac{T_0}{T_{\text{M}}}\right]}\right)
    \end{equation}
\end{subequations}

At a first sight, one can visualize in eq. \eqref{angle_a} that propagation below the XZ plane, i.e., negative angles $\theta_{mn}^{(i)}$, is allowed. A negative transverse wavenumber $k_{m}$ (with real-valued $\beta_{mn}^{(i)}$) is easily achievable for some negative indexes $m$, which translates into a negative diffraction angle $\theta_{mn}^{(i)}$. This constitutes a major difference with respect to our previous works on time-only metasurfaces, where negatives diffraction angles were forbidden \cite{alex2023diffraction}. Additionally, eq. \eqref{angle_b}  reveals that most of the diffraction harmonics tend to close to the normal ($\theta_{mn}\rightarrow0$) when the temporal period $T_{\text{M}}$ decreases or, analogously, when the time-modulation frequency $\omega_{\text{M}}$ increases, with the exception of the $(m, 0)$-th harmonics that are not affected by the time modulation. Therefore, the diffraction pattern produced here is of a much richer nature. Thus, the beamforming capabilities of the present space-time metasurface are expected to significantly exceed those of the previous space-only and time-only configurations.

\subsection{Reflection/Transmission Coefficients}\label{2B}

The derivation of the scattering parameters, i.e., the reflection and transmission coefficients, require of \emph{prior knowledge} of the electric-field profiles at the space-time discontinuity.  Formally, the space-time profile, also referred to as \emph{basis function} (bf), can be described by the function $\mathbf{E}_\mathrm{bf}(y,t)$ depending on space and time. 

The application of the continuity of the electric fields through the space-time interface ($z=0$), 
\begin{equation}\label{E_yt}
    \mathbf{E}_t^{\text{TE/TM},(1)}(y, 0, t) = \mathbf{E}_\mathrm{bf}(y,t) = \mathbf{E}_t^{\text{TE/TM},(2)}(y, 0, t)  \,,
\end{equation}
leads to
\begin{align}
1+R &= T \\
\label{orden0}
1 + R &= \frac{\widetilde{E}_\mathrm{bf}(k_{0}, \omega_0)}{PT_{\text{M}}}  \\
\label{ordenn} 
E_{mn}^{(1)} &= E_{mn}^{(2)} = \frac{\widetilde{E}_\mathrm{bf}(k_{m}, \omega_n)}{PT_{\text{M}}} \, ,
\end{align}
with
\begin{equation}
    \widetilde{E}_\mathrm{bf}(k_{m}, \omega_n) = \displaystyle\int_{-\frac{P}{2}} ^{\frac{P}{2}} \displaystyle\int_{0} ^{T_{\text{M}}} E_\mathrm{bf}(y,t)\, \text{e}^{- \text{j} (\omega_{n} t -  k_{m} y)} \text{d}y\, \text{d}t\, .
\end{equation}
The term ${E}_\mathrm{bf}(y,t)$ is simply the instantaneous amplitude of the vector $\mathbf{E}_\mathrm{bf}(y,t)$ representing the base function and $\widetilde{E}_\mathrm{bf}(k_{m},\omega_{n})$ is its Fourier transform. 

The imposition of the boundary conditions in eq. \eqref{E_yt} presents a fundamental difference with respect to other analytical methods that are also based on the use of Bloch-Floquet series \cite{tiukuvaara2021floquet}. Other methods, after expanding the fields in Bloch-Floquet series to the left (1) and right (2) sides of the space-time interface, directly impose continuity between them, i.e., $\mathbf{E}_t^{\text{TE/TM},(1)} = \mathbf{E}_t^{\text{TE/TM},(2)}$. Then, the fields are projected on each Floquet harmonic to reach a linear system of equations with complex values that gives the Floquet coefficients $E_{mn}$.  This technique share similarities with mode-matching methods. In our case, the approach is different. We project separately the fields of the left and right sides of the space-time discontinuity onto a single basis function $E_\mathrm{bf}$ of our choice, i.e., $\mathbf{E}_t^{\text{TE/TM},(1)} = \mathbf{E}_\mathrm{bf}$ and $\mathbf{E}_t^{\text{TE/TM},(2)} = \mathbf{E}_\mathrm{bf}$. In this case, instead of reaching a complex-valued linear system of equations, we directly obtain the value of the Floquet coefficients $E_{mn}$. This approximation has been successfully applied in the past for the analysis of 1D-spatial \cite{rodriguez2012analytical, rodriguez2012quasi}, 1D-time \cite{alex2023diffraction, moreno2023time}, 2D-spatial \cite{dubrovka2006equivalent, rodriguez2015analytical, molero2021metamaterial} and 3D-spatial \cite{alex2023analytical} periodic systems. It significantly reduces the computation time by avoiding solving the system of equations and allows to extract the associated equivalent circuit shown in Fig. \ref{Admitancias}(a).

 We project separately the fields of the left and right sides of the space-time discontinuity onto a single basis function $E_\mathrm{bf}$ of our choice, i.e., $E^{(1)} = E_\mathrm{bf}$ and $E^{(2)} = E_\mathrm{bf}$ [see eq. (16)].

From eqs. \eqref{orden0} and \eqref{ordenn}, we may also obtain that
\begin{equation}\label{relation}
    E_{mn}^{(1)} = E_{mn}^{(2)} = (1 + R) N(k_{m}, \omega_n) \,,
\end{equation}
where
\begin{equation}\label{Nn}
N(k_{m},\omega_{n}) = \frac{\widetilde{E}_\mathrm{bf}(k_{m}, \omega_n)}{\widetilde{E}_\mathrm{bf}( k_{0}, \omega_0)}    
\end{equation}
is a term that accounts for the coupling between the fundamental $00$-th harmonic and the corresponding $(m,n)$-th space-time harmonic.

Now, the continuity of the instantaneous Poynting vector is imposed at the space-time interface ($z=0$). The power passing through the interface is evaluated over a temporal period $T_{\text{M}}$ and over a spatial period $P$. This leads to 
\begin{multline} \label{Poynting1}
\displaystyle\int_{-\frac{P}{2}} ^{\frac{P}{2}} \displaystyle\int_{0} ^{T_{\text{M}}}\mathbf{E}_\mathrm{bf}(y,t) \times \mathbf{H_t}^{\text{TE/TM},(1)}(y, 0, t)\, \text{d}y\, \text{d}t \\ = \displaystyle\int_{-\frac{P}{2}} ^{\frac{P}{2}} \displaystyle\int_{0} ^{T_{\text{M}}} \mathbf{E}_\mathrm{bf}(y,t) \times \mathbf{H_t}^{\text{TE/TM},(2)}(y, 0, t)\, \text{d}y\, \text{d}t \, 
\end{multline}
By replacing the values of eqs. \eqref{HTE1}, \eqref{HTE2} (TE incidence) or eq. \eqref{HTM} (TM incidence) and eq. \eqref{relation} into eq. \eqref{Poynting1} and operating, we achieve:
\begin{multline}\label{Poynting2}
    (1 -R)Y_{00}^{\text{TE/TM},(1)} - (1 + R) \displaystyle\sum_{\substack{mn \neq 00}}  Y_{mn}^{\text{TE/TM},(1)} \, |N(k_{m},\omega_{n})|^{2}   \\      
    = (1 + R)Y_{00}^{\text{TE/TM},(2)}  + (1 + R) \displaystyle\sum_{\substack{mn \neq 00}}Y_{mn}^{\text{TE/TM},(2)} |N(k_{m},\omega_{n})|^{2}\, ,\\
\end{multline}
 
\noindent By manipulating \eqref{Poynting2}, the reflection coefficient $R$ admits to be expressed as
\begin{equation}\label{R}
    R = \frac{Y_{00}^{\text{TE/TM},(1)} - Y_{00}^{\text{TE/TM},(2)} - Y_\mathrm{eq}}{Y_{00}^{\text{TE/TM},(1)} + Y_{00}^{\text{TE/TM},(2)} + Y_\mathrm{eq}} \,.
\end{equation}
The term $ Y_\mathrm{eq}$, which  groups the double sums in eq. \eqref{Poynting2}, is the \emph{equivalent admittance} that models the space-time metasurface. It is computed as
\begin{equation} \label{Yeq}
    Y_\mathrm{eq} = \displaystyle\sum_{\substack{\forall mn \neq 00}} \left|N(k_{m},\omega_{n})\right|^{2} \left(Y_{mn}^{\text{TE/TM},(1)} + Y_{mn}^{\text{TE/TM},(2)}\right).
\end{equation}

\subsection{Circuit Topology}
\begin{figure*}[!t]
    \subfigure{\includegraphics[width= 2\columnwidth]{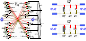}}
	\caption{Circuit topology that describes the physics of the space-time-varying metasurface: (a) Complete equivalent circuit including the two-dimensional sum of infinite transmission lines representing each $(m,n)$-th harmonic. (b) Reduced equivalent circuit for TE and TM incidences.}
	\label{Admitancias}
\end{figure*}

Eqs. \eqref{R} and \eqref{Yeq} are circuitally interpreted by the topology shown in \Fig{Admitancias}. In general, eq. \eqref{R} reveals that the circuit model associated to the analytical framework consists of two semi-infinite transmission lines modeling the input (1) and output (2) media, and an equivalent admittance $Y_\mathrm{eq}$ that models the space-time metasurface [see \Fig{Admitancias}(b)]. A deeper insight into eq. \eqref{Yeq} shows that the equivalent admittance is internally formed by a sum of the admittances (parallel connections of semi-infinite transmission lines) associated to each space-time Floquet harmonic. 

Each of the parameters $N(k_{m},\omega_{n})$ taking part in eq. \eqref{Yeq} is interpreted as a complex transformer that takes into consideration the coupling between the $(m,n)$-th harmonic and the fundamental ($0,0$)th one. Each admittance of each Floquet harmonic is loaded with a complex transformer, except for the fundamental harmonic whose value is $N(k_{0},\omega_{0}) = 1$. This is sketched in \Fig{Admitancias}(a). Note that, for the sake of visualization, only a few harmonics of the double infinite sum are represented  in \Fig{Admitancias}(a), and that the transmission lines of the fundamental harmonic are marked in blue.  

At a first sight, the circuit topologies shown in Figs. \ref{Admitancias}(a) and \ref{Admitancias}(b) coincide with those shown in previous works on space-only \cite{mesa2018unlocking} and time-only \cite{alex2023diffraction} metamaterial configurations. However, the value of the forming admittances and complex transformers, which rule the diffraction behavior of the metasurface, are completely different. In (three-dimensional) space-only systems, the complex transformers are a function of the transverse wavenumbers $k_{m}, k_{n}$ and $k_{l}$. In time-only problems, those are a function of the angular frequency $\omega_{n}$. In spatiotemporal systems, the complex transformers are a function of the transverse wavenumbers plus the angular frequency. 

It was shown that higher-order harmonics in space-only and time-only metasurfaces described by the present topology are of evanescent and propagative nature, respectively. This causes that higher-order harmonics contribute with capacitive and/or inductive terms in space-only (patch-based and aperture-based) metasurfaces \cite{mesa2018unlocking}, while  higher-order harmonics contribute with a purely resistive term in time-only configurations \cite{alex2023diffraction}. The scenario is rather more complex in the present space-time metasurface. Higher-order harmonics   fulfilling $|n|\gg|m|$ are predominantly temporal. Their propagation constant is frequency-independent and real-valued.  It can be approximated as
\begin{equation}
    \beta_{mn}^{(i)}\left|_{|n|\gg|m|} \right. \approx \left|\frac{2\pi n}{cT_\text{M}}\right|
    \sqrt{\varepsilon_r^{(i)}\mu_r^{(i)}}  \, .
\end{equation}
By using eqs. \eqref{admittanceTM} and \eqref{admittanceTE}, the admittances of the $(m,n)$-th harmonic read in this case:
\begin{equation}
    Y_{mn}^{\text{TM/TE}, (i)}\left|_{|n|\gg|m|} \right. \approx Y_{00}\, \sqrt{\frac{\varepsilon_r^{(i)}}{\mu_r^{(i)}}}
\end{equation}
Therefore, all the space-time harmonics with $|n|\gg|m|$ contribute together with a pure resistive term. 

On the other hand, higher-order harmonics fulfilling \linebreak $|m|\gg|n|$ are predominantly spatial. Their propagation constant is frequency-independent and imaginary:
\begin{equation}
    \beta_{mn}^{(i)}\left|_{|m|\gg|n|} \right. \approx -\text{j} \frac{2\pi }{P} \left| m  \right|    \, .
\end{equation}
The associated admittances read:
\begin{subequations}
\begin{equation}
    Y_{mn}^{\text{TM}, (i)}\left|_{|m|\gg|n|} \right. \approx  \text{j} (\omega_0 + n\omega_\text{M})\frac{\varepsilon_r^{(i)} \varepsilon_0 P}{2\pi |m|}
\end{equation}
\begin{equation}
    Y_{mn}^{\text{TE}, (i)}\left|_{|m|\gg|n|} \right. \approx \frac{1}{j (\omega_0 + n\omega_\text{M})}\, \frac{2 \pi |m|}{\mu_r^{(i)} \mu_0 P}
\end{equation}
\end{subequations}
where $\omega_\text{M} = 2\pi / T_\text{M} $ is simply the angular frequency of the time modulation.

From the former expressions, it can be identified that the $(m,n)$-th higher-order TM harmonic fulfilling $|m|\gg|n|$ contributes with \emph{two parallel capacitors}. The first capacitor is dependent on the frequency of the incident wave $\omega_0$ and second one on the time modulation of the space-time metasurface $\omega_\text{M}$. When considering higher-order terms ($|n|\gg 1$), the latter capacitor is the dominant, being the time modulation of the metasurface truly relevant in this case. Similarly, the $(m,n)$-th higher-order TE harmonic fulfilling $|m|\gg|n|$ contributes with \emph{two series inductors}, the one associated to the frequency $\omega_\text{M}$ being the dominant. Therefore, space-time harmonics with $|m|\gg|n|$ contribute jointly with a pure capacitive or inductive term, depending on the type of wave incidence.

The propagation constant of higher-order harmonics that fulfill $|m|\sim|n|$ ($|m|, |n| \gg 1$) is still frequency-independent, but cannot lead to a simple classification of the harmonic contributions into purely resistive or purely capacitive/inductive. Depending on the surrounding media ($\varepsilon_r$ and $\mu_r$) and electrical parameters of the space-time modulation ($P$ and $T_\text{M}$), the propagation constant $\beta_{mn}^{(i)}$ will be real or imaginary and the $(m,n)$-th higher-order harmonic fulfilling $|m|\sim|n|$ will either contribute with a resistive or a capacitive/inductive term.

Conversely, low-order harmonics have associated a propagation constant that is frequency-dependent and a function of the incident angle $\theta_0$, among other parameters. As a consequence, low-order terms cannot be decoupled into purely spatial or purely temporal. Thus, they cannot be directly associated to resistors, capacitors or inductors. To maintain the wideband behavior of the analytical approach, low-order harmonics should be analyzed  directly with their \emph{complex} admittances $Y_{mn}^{\text{TE/TM},(i)}$. Therefore, the whole equivalent admittance $Y_\text{eq}$ in eq. \eqref{Yeq} can be divided into three main parallel contributions: a complex and frequency-dependent admittance $Y_\text{eq}^{\text{(lo)}}$ describing the effect of low-order harmonics, a resistor $R^{\text{(hi)}}$ modeling predominantly-temporal higher-order TM/TE harmonics, and either a capacitor $C^{\text{(hi)}}$ (TM incidence) or an inductor $L^{\text{(hi)}}$ (TE incidence) modeling predominantly-spatial higher-order harmonics. 

If needed, further approximations can be made in the quasi-static regime (low frequencies: $\omega_0\ll 2\pi/T_\text{M}$ and $\omega_0 / c \ll 2\pi/P$) in order to neutralize the frequency-dependence of the complex low-order admittance  $ Y_\text{eq}^{\text{lo}}$. However, this comes at the expense of reducing the operation bandwidth of the wideband analytical model.  In the quasi-static regime, the time modulation truly marks the behavior of the space-time harmonics, i.e., $\omega_n \approx  n \omega_\text{M}$ and $k_{m} \approx 2\pi m/P$. Thus, the admittance of the $(m,n)$-th harmonic under a quasi-static (qs) approximation is
\begin{subequations}
\begin{equation}
    Y_{mn}^{\text{TM}, (i), qs} \approx  \frac{\varepsilon_r^{(i)} \varepsilon_0\, n\, \omega_\text{M}}{ \sqrt{\varepsilon_r^{(i)} \mu_r^{(i)} (n\, \omega_\text{M})^2 - \left(\frac{2\pi m}{P} \right)^2}}
\end{equation}
\begin{equation}
    Y_{mn}^{\text{TE}, (i), qs} \approx  \frac{\sqrt{\varepsilon_r^{(i)} \mu_r^{(i)} (n\, \omega_\text{M})^2 - \left(\frac{2\pi m}{P} \right)^2}}{\mu_r^{(i)} \mu_0\, n\, \omega_\text{M} }\, ,
\end{equation}
\end{subequations}
which can only be purely real or reactive depending on the selected $(m,n)$-th harmonic. As a consequence, the equivalent admittance in eq. \eqref{Yeq} can be reduced in the quasi-static case to a parallel connection of a resistor $R^\text{(hi), qs}$ and a capacitor $C^\text{(hi), qs}$ (TM incidence) or the resistor $R^\text{(hi), qs}$ and an inductor $L^\text{(hi), qs}$ (TE incidence).

\subsection{Basis Functions}

\begin{figure*}[!t]
    \includegraphics[width= 2\columnwidth]{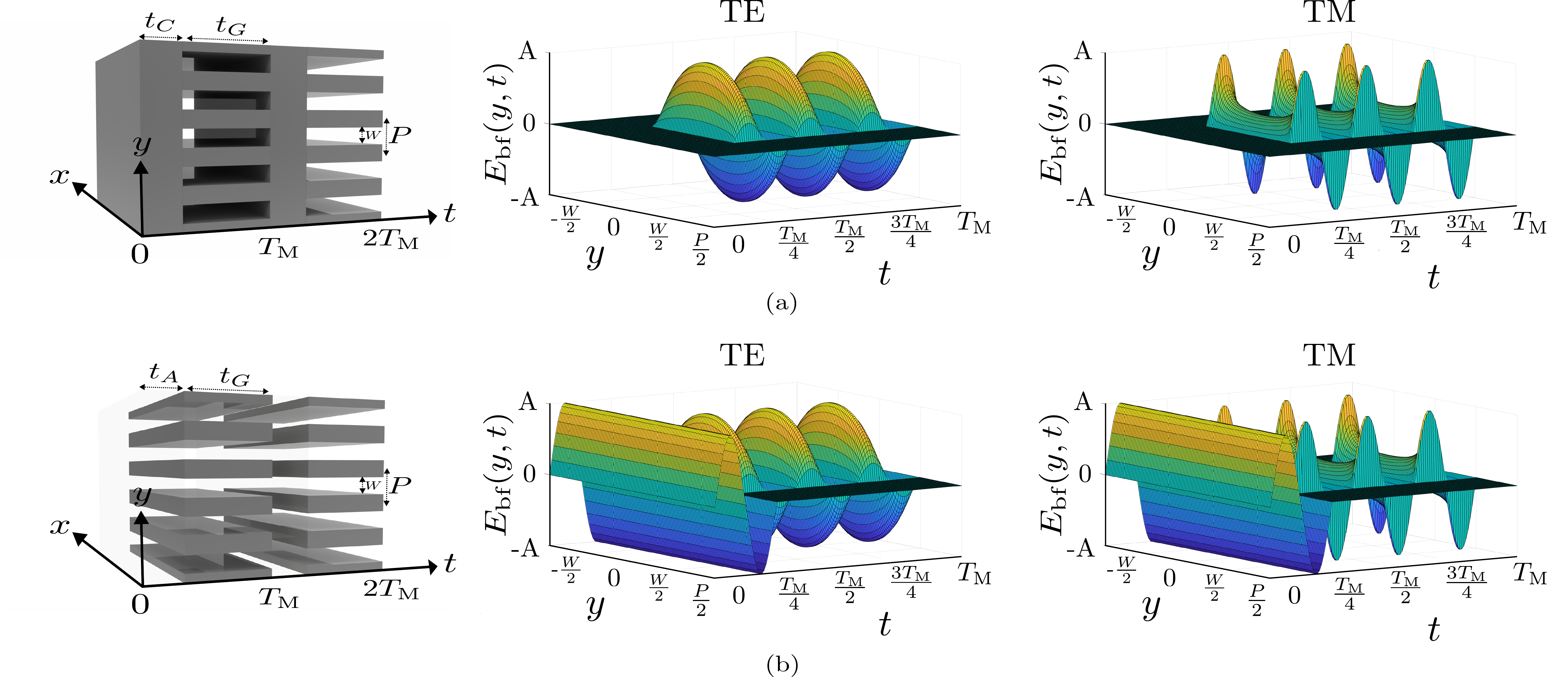}
    \caption{\vspace{-0.0cm}(a) Sketch of the C-G metasurface and the $\mathbf{E}_\mathrm{bf}(y,t)$ on a single space-time unit cell of the C-G configuration under $\text{TE/TM}$ normal incidence. (b) Sketch of the A-G metasurface and the $\mathbf{E}_\mathrm{bf}(y,t)$ on a single space-time unit cell of the A-G configuration under $\text{TE/TM}$ normal incidence. Parameters: $T_{\text{M}}=4T_{\text{0}}$, $t_C = t_A = [0,\ 0.25T_{\text{M}}]$, $t_G = [0.25T_{\text{M}},\ T_{\text{M}}]$, $P=0.7\lambda_0$ and $W=0.5P$.}
	\label{Fig4}
\end{figure*}

As detailed in Section \ref{2B}, the knowledge of the tangential electric field profile (basis function) $\mathbf{E}_\mathrm{bf}(y,t)$ is crucial for the analysis of the space-time metasurface via the analytical method. We expect to obtain accurate results as long as the selected basis functions carry actual physical information of the electromagnetic fields taking place in the considered scenario.
In this subsection, we describe the analytical expressions for the space-time base functions involving the three states under consideration: air (A), conductor (C) and grating (G). 
\subsubsection{Air state (A)}
In the air state, the space-time metasurface \emph{vanishes} from an electromagnetic perspective. Therefore, the base function that models the air state coincides with the electric profile of the incident plane wave; namely,
\begin{equation} \label{base_a}
    E_\mathrm{bf}(y,t)= A \sin(\omega_0 t), \, \, \forall y \in \bigg[-\frac{P}{2}, \frac{P}{2}\bigg], t\in t_A,
\end{equation} 
where $t_A$ is the time in which the space-time metasurface is in the air state.
\subsubsection{Conductor state (C)}
In the conductor state, the space-time metasurface turns into an infinitesimally-thin metallic sheet. If losses are neglected, the metallic sheet can be modeled with PEC conditions, so the tangential electric field must be zero at its surface. This leads to the base function
\begin{equation} \label{base_c}
    E_\mathrm{bf}(y,t)= 0, \, \,\forall y \in \bigg[-\frac{P}{2}, \frac{P}{2}\bigg], t\in t_C,
\end{equation} 
where $t_C$ is the time in which the metasurface is in the conductor state.
\subsubsection{Grating (G)}
The grating state can be modeled with the following base function
\begin{equation} \label{base_g}
    E_\text{bf}(y,t)  = A \sin(\omega_0 t)
    \begin{cases}
         \Bigg[1-\bigg(\frac{2y}{W}\bigg)^{2}\Bigg]^{\frac{1}{2}}, \quad &\text{TE pol.}\\
         \Bigg[1-\bigg(\frac{2y}{W}\bigg)^{2}\Bigg]^{-\frac{1}{2}}, \quad &\text{TM pol.}\, ,\\
    \end{cases}
\end{equation}
defined in the slit region ($-W/2 \leq y \leq W/2$) and the time interval $t\in t_G$, time in which the space-time metasurface is in the grating state. This base function relates the field excitation and charge distribution in a conventional one-dimensional diffraction grating illuminated by a sinusoidal plane wave of frequency $\omega_0$.

As previously introduced, a wise combination (A-G, C-G) of the three states in a \emph{time-periodic scheme} can lead to a rich diffraction phenomenology from which  beamformers can benefit. Left side of \Fig{Fig4}(a) sketches the scenario where the space-time metasurface alternates between conductor and grating states (C-G case). As shown, the scheme is periodic in space and time, with periods $P$ and $T_\mathrm{M}$, respectively. Moreover, the terms $t_C$ and $t_G$ indicate the time in which the metasurface is in the conductor and grating states. Note that $T_\mathrm{M} = t_C + t_G$ in this case to retain time periodicity. The basis function that models the C-G scenario is plotted at the rigth side  of \Fig{Fig4}(a) for TE and TM incident polarizations. Null regions indicate that the space-time metasurface is in the conductor state [eq. \eqref{base_c}], while $\cap$- and $\cup$-shaped regions [eq. \eqref{base_g}] are associated to the grating state under TE and TM incidences, respectively.

Similar rationale is applied to the air-conductor (A-G) case, sketched at left side of \Fig{Fig4}(b). The A-G configuration is periodic in space and time, with periods $P$ and $T_\mathrm{M}$, respectively. In this case, the time period is constituted by two addends, $T_\mathrm{M} = t_A + t_G$, related to the time in which the metasurface is in the air and grating states. The basis function of the A-G case is illustrated in the right side of \Fig{Fig4}(b). $\cap$/$\cup$-shaped regions indicate that the space-time metasurface is in grating state, while sinusoidal regions indicate that the metasurface is in the air state [eq. \eqref{base_a}].

The different states of the structure can also be invoked by regarding  reconfigurable metasurfaces, which for this case would consist of a reconfigurable grating. Reconfiguration is realized via active elements such as diodes or varactors \cite{zhang2018space}, or via electronic materials such as graphene \cite{molero2021metamaterial} or $\text{Ge}_{2}\text{Sb}_{2}\text{Te}_{2}$ \cite{li2022intelligent}. Their electronic effect over the grating is tuned to vary periodically, invoking periodic cycles in which the states of full transparency (A), full reflectivity (C), or the natural grating (G) interchange in the sense described above. This can be possible thanks to the excitation of resonant states in the grating through these elements. The existence of a transient regime when passing from one state to the other is neglected since we are assuming transient time to be temporarily short enough in comparison with the rest of times ($t_{C}, t_{A}, t_{G}$) taking part in a periodic cycle. Small transient times have negligible effects on the Fourier Transform of the electric field profile at the discontinuity \cite{wu2020space}.

\subsection{On the Limitations of the Analytical Approach}

As we fundamentally work with air and metal states,  we first have to define the properties of these elements. Without loss of generality, the air state is modeled as a dielectric of electrical properties $\varepsilon_r = \mu_r = 1$. A first approximation comes when defining the properties of metal-based states: conductor and grating. We consider here that metals are modeled as \emph{perfect electric conductors} (PEC). The assumption of PEC is expected to give accurate results in radio, microwave and millimeter-wave frequencies, where good conducting metals such as copper, gold or silver behave essentially as PEC. When working with metasurfaces and frequency selective surfaces (FSS), it is a common practice in the literature to approximate the response of metals as perfect electric conductors in the aforementioned frequency ranges \cite{rodriguez2012analytical, dubrovka2006equivalent, alex2021exploring}. In fact, we expect that the PEC assumption give accurate results up to the low-THz regime ($<10$ THz), as long as transients can be neglected in the system.  However, the approximation of PEC to model metal-based states would cease to be physical, and therefore valid, in the optical range and at higher frequencies. 

The considered basis functions also limit the frequency range of the analytical approach.  Of the three states in which the space-time metasurface can be (air, conductor and grating), the grating state is the most limiting from an analytical perspective, but the richest one in terms of diffraction. In \cite{alex2021exploring}, it is reported that the maximum operating frequency to be covered with eq. \eqref{base_g} may go beyond to $f \sim 1.5c / P$. For instance, for two different spatial periods of $P=3$ mm and $P=3$ $\mu$m, the maximum frequencies in the grating state may not overpass 150 GHz and 150 THz, respectively. When oblique incidence is considered, this range could decrease to $f_\mathrm{max} \sim  c/P$ (100 GHz and 100 THz in the former example). Moreover,  the slit width should be less than $W \lesssim 0.7P$ to obtain accurate results \cite{alex2021exploring}. Otherwise, it would be more convenient to directly work with currents instead of with electric field profiles \cite{rodriguez2012analytical}. As seen, the frequency range is wide, but naturally limited, when modeling the grating state. 

The maximum operation frequency $f_\mathrm{max}$ of the analytical approach would be determined by the worst-case scenario of the two main approximations discussed above. If the maximum frequency  is exceeded, the accuracy in the results is expected to degrade significantly. Above $f_\mathrm{max}$, the proposed basis functions could not be representative of the actual physical phenomena occurring in the spatiotemporal metastructure. Additionally, metals could  stop being  fully-reflective objects (PEC scenario) and start being semi-transparent from the incident wave's perspective. Therefore, the actual reflection and transmission coefficients can drastically change with respect to the theoretical estimations.

Moreover, above the maximum operation frequency $f_\mathrm{max}$ of the analytical circuit, frequency dispersion and transients could play an important role in the response of the metastructure. Transient effects, out of the scope of the present manuscript, can completely redefine the actual electromagnetic response of  the space-time metasurface. Transients are expected to be non-negligible at frequencies close to the plasma frequency $\omega_p$ (Drude model).  In good conductors such as copper, gold or silver, the plasma frequency is of the order of $\omega_p \sim 10^{16}$ rad/s, and the relaxation times are $\tau \sim 10^{-14}$ s. Thus, the effect of transients should be appreciable around frequencies of the order of $10^{14}$ rad/s. Below this limit (radio, microwaves, millimeter-waves), transients may be neglected without significantly affecting the response of the system.


\section{Wireless Applications}\label{Resultados1}

In this section, we test the capabilities of the space-time metasurface for wireless applications. We show how the electrical tuning of the different space-time parameters can lead to a rich diffraction phenomenology, mixing frequencies, setting specific angular regions and favoring transmission or reflection. Furthermore, we validate the present analytical approach by comparing the results with full-wave simulations in a self-implemented finite-difference time-domain (FDTD) code.

\subsection{Harmonic Distribution} \label{Harmonics}

\begin{figure}[!t]
	\includegraphics[width= 1\columnwidth]{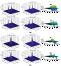}
    \caption{Normalized amplitude of the Floquet coefficients when varying the temporal period in a C-G scenario. Normal TM and TE incidences are considered. Cases: (a) $T_{\text{M}}=8T_0$ and (b) $T_{\text{M}}=2T_0$. The basis function  $E_\mathrm{bf}(y,t)$ is plotted along the results. Parameters: $t_C = [0,\ 0.5T_{\text{M}}]$, $t_G = [0.5T_{\text{M}},\ T_{\text{M}}]$, $P=0.7\lambda_0$, $W=0.5P$.}
    \label{C-G_Tm}
\end{figure}

We start by considering the case in which the space-time metasurface alternates between the conductor and grating states. The basis function that models the C-G case is mathematically described by a combination of eqs. \eqref{base_c} and \eqref{base_g}.  \Fig{C-G_Tm} depicts the obtained normalized amplitudes of the Floquet harmonics under normal TE and TM incidences. In all these simulations, the incidence of a monochromatic wave with a frequency of $30$ GHz has been assumed.  The results are obtained with the proposed analytical model and then compared with our FDTD code. The agreement between the analytical and FDTD results is good, thus validating the present approach. 

 The harmonic pattern shows similarities for TE and TM incidences. Note that two main lobes appear in \Fig{C-G_Tm} for $n=0$ and $n=-16$. These indexes correspond to the frequencies $\omega_n = \pm \omega_0$, respectively. However, in the TM case, the amplitude of the spatial ($m$-indexed) harmonics show a slower decay; namely, a larger number of spatial modes are needed to accurately reconstruct the basis function. This is due to the singularity of the spatial profile present in the TM case that is not in the TE case. On the other hand, the harmonic pattern under TE incidence is mainly governed by the temporal $n$-indexed harmonics.

 \Fig{C-G_Tm}(b) presents the Floquet harmonic pattern in the case of reducing the time period from $T_{\text{M}}=8T_0$ to $T_{\text{M}}=2T_0$. The peaks are observed for the temporal harmonics $n=0$ and $n=-4$. As illustrated, the two main lobes approach each other in the case of reducing the time period (increasing the switching frequency of the metasurface).  The spatial and temporal harmonic dependencies are decoupled, as eqs. \eqref{base_a}-\eqref{base_g} impose and \Fig{C-G_Tm} confirms. Thus,  the behaviour of the spatial modes for both incidences is not affected by modifying the time period.
 
\begin{figure}[t!]	     
    \includegraphics[width=0.9\columnwidth]{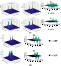}
	\caption{Normalized amplitude of the Floquet coefficients in a C-G scenario.  Normal TM incidence is assumed and the basis function  $E_\mathrm{bf}(y,t)$ is plotted along the results. (a) The duty cycle is varied. First row: $D=0.25$. Second row: $D=0.75$. Here, $D$ and $1-D$ indicates the amount of time that the metasurface is in the conductor and grating states, respectively. Parameters: $T_{\text{M}}=8T_0$, $P=0.7\lambda_0$ and $W=0.5P$.  (b) The spatial ratio $W/P$ is varied.  First row: $W=0.1P$. Second row: $W=0.5P$. Parameters: $T_{\text{M}}=4T_0$, 
 $P=0.7\lambda_0$ and  $D=0.5$.  }
	\label{C-G_D_W}
\end{figure}

Another temporal parameter of interest is the \emph{duty cycle} $D$ \cite{moreno2023time}. The duty cycle is a dimensionless parameter that relates the amount of time that the space-time metasurface is in each of the three states. As it will be discussed later, the duty cycle plays an important role in controlling the amplitude of the diffracted, reflected and transmitted, waves. For the present C-G case, the metasurface keeps in conductor state a time given by $DT_\mathrm{M}$, with $T_{\text{M}}$ being the temporal macroperiod. Otherwise, it remains in the grating state $(1-D)T_\mathrm{M}$ seconds. First and second row of \Fig{C-G_D_W}(a) show the harmonic pattern when the duty cycle is $D=0.25$ and $D=0.75$, respectively. Increasing the duty case in this case implies a greater presence of the grating state. Therefore, the width of the main lobes narrows along the temporal ($n$-indexed) axis as the spatiotemporal field profile (basis function) is more similar to that of a pure sine.

The last spatial parameter of interest is the relation between the slit width $W$ and the spatial period $P$; namely, the dimensionless ratio $W/P$. Logically, decreasing the slit width in a grating implies an increase in the reflection through the structure. From a harmonic perspective [see the harmonic pattern in \Fig{C-G_D_W}(b)], decreasing the ratio $W/P$ provokes that the width of the main lobes widens. This is specially noticeable in the TM case, as the basis function includes the presence of two singularities at the edge of the slits (positions $y = \pm W/2$ within the unit cell) in a $\cup$-shaped profile. The two singularities get closer as $W/P$ reduces. As a consequence, more spatial modes are needed to reconstruct the tangential electric-field profile due to the rapid spatial variation of the basis function along $y$.

\subsection{Diffracted Waves}
\begin{figure}[!t]
	\centering
	\includegraphics[width= 1\columnwidth]{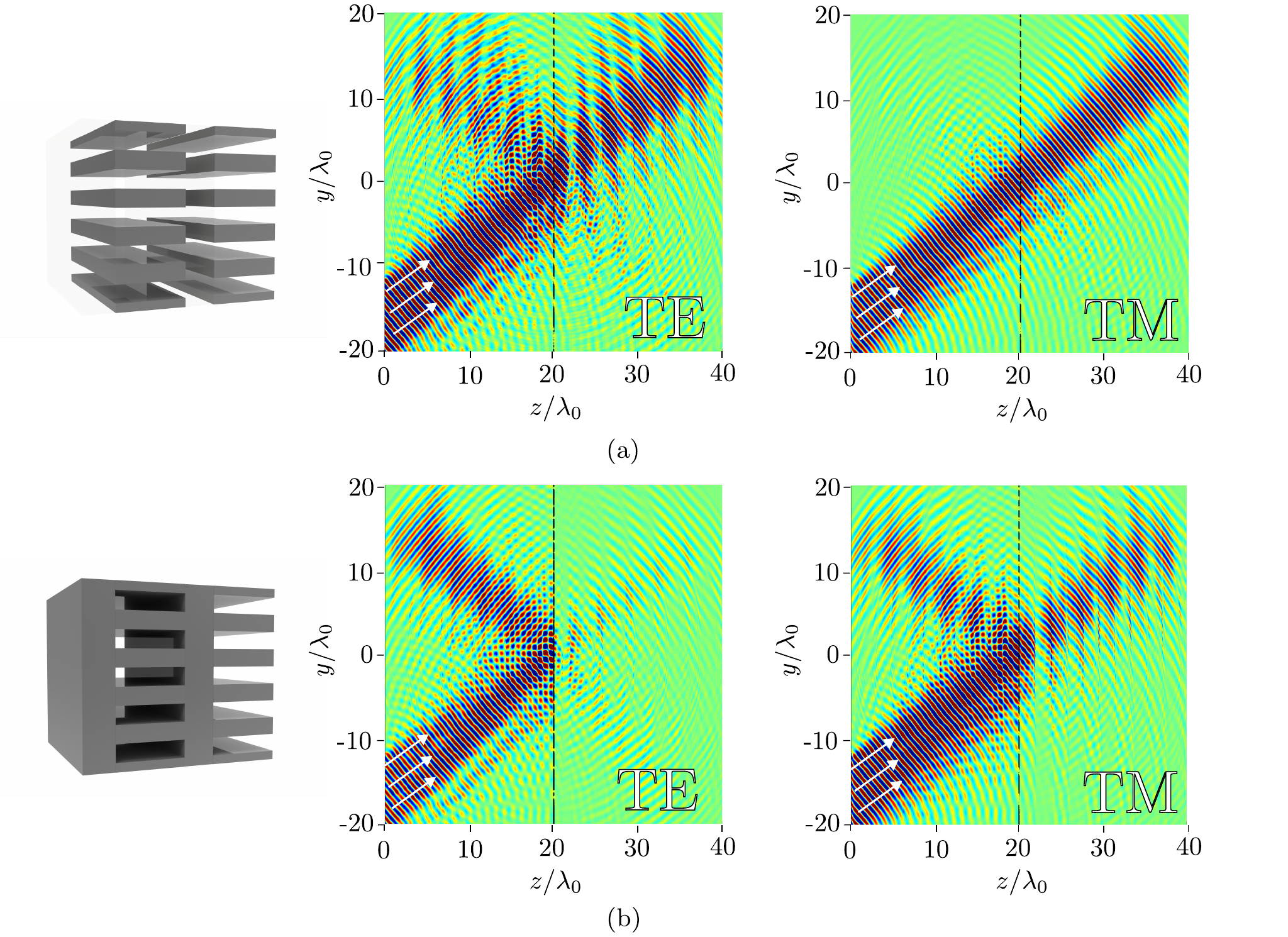}
	\caption{FDTD simulation of the diffracted electric fields for the (a) A-G and the (b) C-G configurations. The space-time metasurface is placed at \linebreak $z/\lambda_0=20$. Oblique TE and TM incidences ($\theta_0=40^{\text{o}}$) are analyzed. Parameters:  $T_{\text{M}}=4T_0$, $t_C = t_A = [0,\ 0.5T_{\text{M}}]$, $t_G = [0.5T_{\text{M}},\ T_{\text{M}}]$, $P=0.7\lambda_0$, $W=0.5P$. }
	\label{R_vs_T}
\end{figure}
To gain a more comprehensive understanding on the diffraction patterns of these metasurfaces, \Fig{R_vs_T}(a) and \Fig{R_vs_T}(b) illustrate the electric-field distributions of the A-G and C-G cases, respectively, extracted with the self-implemented FDTD code. The same duty cycle $D=0.5$ is maintained in both configurations for a fair comparison. 

In general, higher transmission is observed for TM polarization than for TE polarization. This is in line with the results obtained in previous works on spatially-modulated surfaces \cite{rodriguez2012analytical,rodriguez2012quasi}. However, when the metasurface alternates between air and grating states [A-G configuration in \Fig{R_vs_T}(a)], a high degree of transmission is generally seen. In contrast, when the metasurface switches between the conductor and grating states  [C-G configuration in \Fig{R_vs_T}(b)], high reflection is created. This is a somewhat expected but interesting feature from an engineering perspective, since the air and conductor states tend to favor transmission and reflection, respectively.  This feature of the space-time metasurface, combined with the tuning of the duty cycle parameter, can be exploited in practice for the efficient design of reflection-based, transmission-based or mixed reflection-transmission-based beamformers.
 
From a physical perspective, it is noteworthy that the incorporation of spatial modulation by the grating leads to the emergence of spatial modes in the upper ($y\geq0$) and lower ($y<0$) half-spaces. This constitutes a major difference with respect to our previous works on time-only modulated metamaterials, where only diffraction in the upper half-space was permitted \cite{alex2023diffraction, moreno2023time}. Thus, the present space-time metasurface creates a richer diffraction pattern compared to time-only configurations. 

\subsection{Frequency Mixing} \label{F_mixing}
\begin{figure}[!t]
    \centering
	\subfigure{\includegraphics[width= 1\columnwidth]{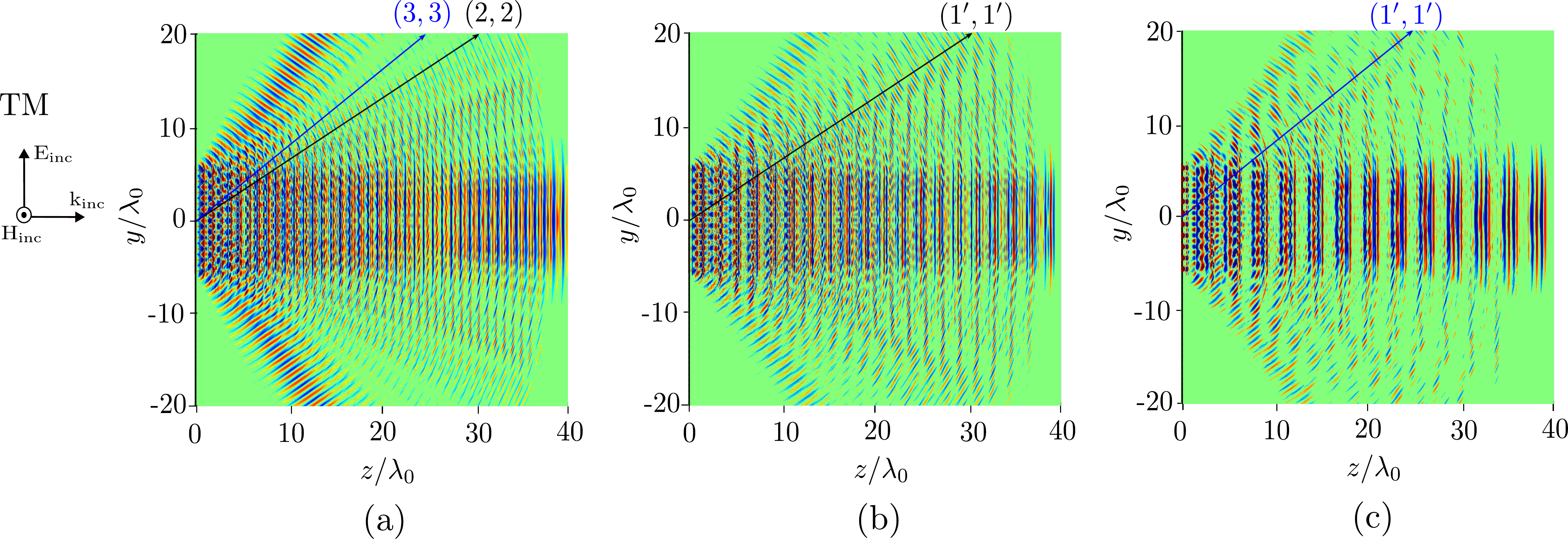}}
	\caption{FDTD results for a C-G frequency mixer for an incident wave of frequency $\omega_0$ = $2\pi \cdot 30$ GHz: (a) $T_{\text{M}}=T_0$; (b) $T'_{\text{M}}=2T_0$;  (c) $T'_{\text{M}}=3T_0$. Normal TM incidence is considered. Parameters: $t_C = [0,\ 0.25T_{\text{M}}]$, $t_G = [0.25T_{\text{M}},\ T_{\text{M}}]$, $P=1.2\lambda_0$, $W=0.5P$.}
	\label{mixer}
\end{figure}

Frequency mixing is vital for some physics and engineering applications such as radio astronomy, telecommunications or radar systems. Traditionally, nonlinear components such as Schottky diodes, GaAs FETs and CMOS transistors are used for this purpose. In this regard, temporal and spatiotemporal metamaterials possess the inherent ability to mix frequencies \cite{taravati2018aperiodic}. This is essentially due to ``temporal" diffraction and the production of higher-order Floquet harmonics ruled by the condition $\omega_{n} = \omega_{0} + 2\pi n / T_{\text{M}}$ [see eq. \eqref{wn}]. Thus, the proposed space-time metasurface can produce new frequencies, $\omega_n$, from the frequency of the incident wave, $\omega_0$.

Interestingly, eq. \eqref{angle_a} suggests that two different space-time modulations can lead to the same diffraction angle, but with different output frequencies, by simply tuning their space-time parameters. In order to show this frequency-mixing effect, let us consider the two following modulations. Modulation \#1 is defined by the (unprimed) space-time parameters $\{m,n, P, T_\text{M} \}$. Similarly, modulation \#2 is defined by the (primed) parameters $\{m',n', P', T'_\text{M} \}$. This leads to the diffraction angles $\theta_{mn}$ and $\theta_{m'n'}$ (associated to modulations \#1 and \#2, respectively), given by
\begin{equation} \label{angle_mn}
   \theta_{mn} = \text{asin} \left( \frac{k_t + \frac{2\pi m}{P}}
    {\frac{2\pi}{cT_0} \sqrt{\varepsilon_r^{(1)} \mu_r^{(1)}} \Big(1 + n \frac{T_0}{T_\text{M}} \Big)} \right) 
\end{equation}
\begin{equation} \label{angle_mn_prime}
   \theta_{m'n'} = \text{asin} \left( \frac{k_t + \frac{2\pi m'}{P'}}
    {\frac{2\pi}{cT_0} \sqrt{\varepsilon_r^{(1)} \mu_r^{(1)}} \Big(1 + n' \frac{T_0}{T'_\text{M}} \Big)} \right) 
\end{equation}
Nonetheless, both configurations are expected to have the same spatial period ($P' = P$), since this parameter cannot be electronically reconfigured in practice like the duty cycle $D$ or the time period $T_\text{M}$. Under this assumption, by equating eqs. \eqref{angle_mn} and \eqref{angle_mn_prime},   we have  
\begin{equation} \label{mixing1}
    \frac{k_t + \frac{2\pi m}{P}}
    {\frac{2\pi}{cT_0} \sqrt{\varepsilon_r^{(1)} \mu_r^{(1)}} \Big(1 + n \frac{T_0}{T_\text{M}} \Big)}
    =
    \frac{k_t + \frac{2\pi m'}{P}}
    {\frac{2\pi}{cT_0} \sqrt{\varepsilon_r^{(1)} \mu_r^{(1)}} \Big(1 + n' \frac{T_0}{T_\text{M}'} \Big)}
\end{equation}
The former equation reduces under normal incidence ($k_t = 0$) to
\begin{equation} \label{mixing2}
    \frac{m}{m'} = \frac{1 + n \frac{T_0}{T_\text{M}}}{1 + n' \frac{T_0}{T_\text{M}'}}
\end{equation}

Expressions \eqref{mixing1} and \eqref{mixing2} give us the design equations to produce frequency mixing between harmonics of different orders ($\{m,n\}$ and $\{m',n'\}$) while maintaining the same output diffraction angle. This can be done by simply adjusting the temporal periods of the two modulations, $T_\text{M}$ and $T'_\text{M}$. 

\Fig{mixer} shows a practical example of frequency mixing. For simplicity, we have selected an scenario ruled by normally-incident waves. However, the previous discussion can be straightforwardly extended to the design of spatiotemporal frequency mixers that operate under oblique-incidence conditions by using eq. \eqref{mixing1}. In the figure, a monochromatic plane wave of frequency $\omega_0$ impinges normally on a C-G space-time metasurface (located at $z/\lambda_0=0$). As seen, the metasurface creates a rich diffraction pattern, where only the transmitted diffracted fields are plotted for a better visualization of the results. With the design information given by eq. \eqref{mixing2}, the temporal periods of \Figs{mixer}(a), \ref{mixer}(b), \ref{mixer}(c) have been carefully chosen. The underlying idea is to keep the same diffraction angle while varying the output frequencies in the three configurations that are shown. For instance, it can be seen that the diffraction orders ($m=2, n=2$) and ($m' = 1, n'=1$) in \Figs{mixer}(a) and \ref{mixer}(b) share the same diffraction angle ($\theta_{22} = \theta_{1'1'} = 33.72^\mathrm{o}$) while their output frequencies are different ($\omega_{2}=3\omega_0$, $\omega_{1'} = 1.5\omega_0$). The aforementioned diffraction orders are marked with a blue line in \Figs{mixer}(a) and \ref{mixer}(b). Similarly, the diffraction orders ($m=3, n=3$) and ($m' = 1, n'=1$), marked with a black line in \Figs{mixer}(a) and \ref{mixer}(c), have the same diffraction angle ($\theta_{33} = \theta_{1'1'} = 38.65^\mathrm{o}$) while their output frequencies are $\omega_{3}=4\omega_0$ and $\omega_{1'} = 1.33\omega_0$, respectively. As this example seeks to illustrate, the capabilities of the present metal-based metasurface are promising for the production of frequency mixers and frequency multipliers based on space-time architectures that could be integrated in wireless communication systems.

\subsection{Transmissive Beamformer}
\begin{figure*}[!t]
	\centering
    \includegraphics[width= 2\columnwidth]{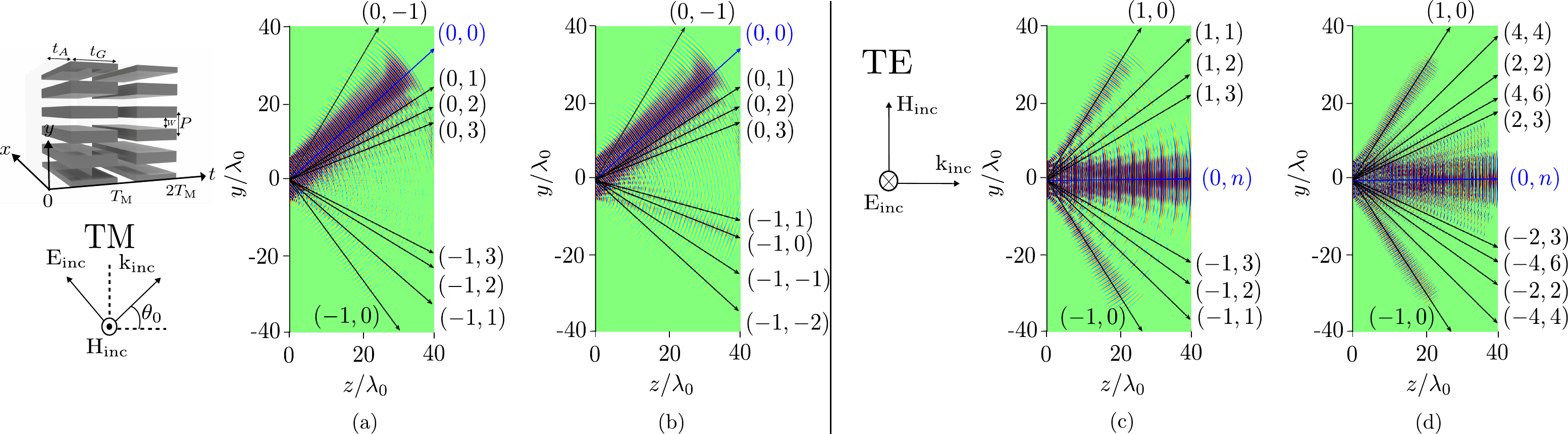}
	\caption{(a)-(b) Diffraction patterns of the A-G configuration under oblique TM incidence ($\theta_0=40^\mathrm{o}$): (a) $T_{\text{M}}=4T_{\text{0}}$ and $P=0.7\lambda_0$. (b) Different spatial period: $P=\lambda_0$. Fixed parameters: $t_A = [0,\ 0.25T_{\text{M}}]$, $t_G = [0.25T_{\text{M}},\ T_{\text{M}}]$, $W=0.5P$. (c)-(d) Diffraction patterns of the A-G configuration under normal TE incidence: (c) $T_{\text{M}}=4T_{\text{0}}$ and $P=1.2\lambda_0$. (d) Different temporal period: $T_{\text{M}}=T_0$. Fixed parameters: $t_A = [0,\ 0.25T_{\text{M}}]$, $t_G = [0.25T_{\text{M}},\ T_{\text{M}}]$, $W=0.6P$.}
	\label{A-G_beam}
\end{figure*}

The main purpose of this section is to exploit the performance of these structures as transmission beamformers. Thus, we test the A-G configuration by changing the spatial and temporal periods. For a better visualization in the simulations, only the transmissive part of the diagrams is plotted. The A-G metasurface is placed at the beginning of the space simulation ($z/\lambda_0=0$). The most important diffraction orders have been marked with an arrow to facilitate their visualization. and their space and time indexes $(m,n)$ have been included to know their nature. 

Thus, \Fig{A-G_beam}(a) depicts the diffraction pattern of a monochromatic wave that impinges in the spatiotemporal metasurface under oblique TM incidence ($\theta_0=40^{\text{o}}$). Note how, in the upper part of the transmission plane ($y/\lambda_0>0$), the diffraction angles must be similar to the time-only case since the  $(m=0,n)$-th harmonics are implied. However, in this manuscript, the beamformer capabilities have been increased since the arising of the new space-time harmonics in the lower part of the transmission diagram ($y/\lambda_0<0$). In these examples, it is noticeable the emerging of the $(-1,n)$-th harmonics. 

\Fig{A-G_beam}(b) shows the same scenario but with a slight increase of spatial period (from $P=0.7\lambda_0$ to $P=\lambda_0$). The rest of the parameters, $T_{\text{M}}=4T_0$, $W=0.5P$ and $D=0.25$, are kept fixed. \Fig{A-G_beam}(b) illustrates how $(m=-1,n)$-th harmonics approach the normal as the spatial period $P$ and the temporal index $n$ increase.
This is well predicted by eq. \eqref{angle_b}. Moreover, notice how the temporal $(m=0,n)$-th harmonics are not affected by a change in the spatial period, since the temporal period is the same in both \Figs{A-G_beam}(a) and \ref{A-G_beam}(b).

In contrast, \Figs{A-G_beam}(c)-(d) depict the phenomenology when the temporal period is modified. For simplicity, TE normal incidence is assumed and slit width is increased to $W = 0.6P$ in order to achieve a higher transmission. Furthermore, a spatial period greater than the wavelength of the incident wave is imposed to enrich the diffraction diagram since the first (spatial) grating lobes are excited. \Fig{A-G_beam}(c) shows the diffraction pattern for the same time period as considered throughout the previous section of the manuscript ($T_{\text{M}}=4T_0$). In this case, a symmetric pattern is appreciated where the temporal $(0,n)$-th harmonics are in the same direction ($\theta_{0n}=0^\mathrm{o}$).  Moreover, all the $(0,n)$-th harmonics are propagative due to their temporal nature and the normal incidence conditions. However, since spatial modulation has been included, new spatiotemporal harmonics emerge symmetrically at both the upper and lower parts of the diagram. Finally, in \Fig{A-G_beam}(d), it is observed how the higher-order $(m,n)$-th harmonics approach to the normal when the temporal period is reduced (from $T_{\text{M}}=4T_0$ to $T_{\text{M}}=T_0$), as it was predicted in Section \ref{Angulos}.

Table \ref{table1} provides a comparison of the diffraction angles for the previous cases shown in \Fig{A-G_beam}. They have been extracted with the analytical approach [by means of eq. \eqref{angle_b}] and the FDTD method. A good agreement is appreciated  between the analytical and numerical results.

\begin{table}[!t]
\caption{Diffraction angles obtained for each $(m,n)$-th harmonic using the FDTD and analytical method for the cases shown in \Fig{A-G_beam}.}
\centering \label{table1}
\begin{tabular}{c *5c} 
\toprule
 & \multicolumn{1}{c}{Incidence} & ${\theta_{mn}} ({^\mathrm{o}})$ & \text{Analytical}  & \text{FDTD}    \\ 
\midrule
\multirow{13}{1.5cm}{${T_{\text{M}}=}4T_0$ ${P=}0.7\lambda_0$} & & $\theta_{0-1}$ & $58.98$ & $58.17$  \\ 
\cmidrule(l){3-5}
& & $\theta_{00}$  &  $40.00$ &  $40.36$ \\
\cmidrule(l){3-5}
& & $\theta_{01}$ &  $30.94$ &  $31.27$ \\
\cmidrule(l){3-5}
&  & $\theta_{02}$  &  $25.37$ &  $25.40$ \\
\cmidrule(l){3-5}
& ${\theta_{0}^{\text{TM}}=}40^\mathrm{o}$ & $\theta_{03}$  &  $21.54$ &  $20.38$ \\
\cmidrule(l){3-5}
& & $\theta_{-13}$ & $-26.64$ & $-25.98$  \\ 
\cmidrule(l){3-5}
& & $\theta_{-12}$ &  $-31.54$ &  $-30.20$ \\
\cmidrule(l){3-5}
&  & $\theta_{-1-1}$  &  $-38.89$ &  $-39.43$ \\
\cmidrule(l){3-5}
&  & $\theta_{-10}$  &  $-51.70$ &  $-52.40$ \\
\midrule
\multirow{13}{1.5cm}{${T_{\text{M}}=}4T_0$ ${P=}\lambda_0$} & & $\theta_{0-1}$ & $58.98$ & $58.17$  \\ 
\cmidrule(l){3-5}
& & $\theta_{00}$  &  $40.00$ &  $40.36$ \\
\cmidrule(l){3-5}
& & $\theta_{01}$ &  $30.94$ &  $31.27$ \\
\cmidrule(l){3-5}
&  & $\theta_{02}$  &  $25.37$ &  $25.40$ \\
\cmidrule(l){3-5}
& ${\theta_{0}^{\text{TM}}=}40^\mathrm{o}$ & $\theta_{03}$  &  $21.54$ &  $20.38$ \\
\cmidrule(l){3-5}
& & $\theta_{-11}$ & $-16.57$ & $-15.28$  \\ 
\cmidrule(l){3-5}
& & $\theta_{-10}$ &  $-20.88$ &  $-21.18$ \\
\cmidrule(l){3-5}
&  & $\theta_{-1-1}$  &  $-28.38$ &  $-31.38$ \\
\cmidrule(l){3-5}
&  & $\theta_{-1-2}$  &  $-45.48$ &  $-41.50$ \\
\midrule
\multirow{6}{1.5cm}{${T_{\text{M}}=}4T_0$ ${P=}1.2\lambda_0$} & & $\theta_{\pm10}$ & $\pm56.38$ & $\pm56.37$  \\ 
\cmidrule(l){3-5}
& \multirow{3}{1.5cm}{${\theta_{0}^{\text{TE}}=}0^\mathrm{o}$} & $\theta_{\pm11}$  &  $\pm41.77$ &  $\pm42.61$ \\
\cmidrule(l){3-5}
& & $\theta_{\pm12}$ &  $\pm33.72$ &  $\pm34.41$ \\
\cmidrule(l){3-5}
&  & $\theta_{\pm13}$ &  $\pm28.41$ &  $28.7$ \\
\midrule
\multirow{7}{1.5cm}{${T_{\text{M}}=}T_0$ ${P=}1.2\lambda_0$} & & $\theta_{\pm10}$ & $\pm56.38$ & $\pm55.98$  \\ 
\cmidrule(l){3-5}
& & $\theta_{\pm44}$  &  $\pm41.77$ &  $\pm43.15$ \\
\cmidrule(l){3-5}
& ${\theta_{0}^{\text{TE}}=}0^\mathrm{o}$ & $\theta_{\pm22}$ &  $\pm33.72$ &  $\pm34.31$ \\
\cmidrule(l){3-5}
&  & $\theta_{\pm46}$ &  $\pm28.41$ &  $27.99$ \\
\cmidrule(l){3-5}
&  & $\theta_{\pm23}$ &  $\pm24.60$ &  $23.82$ \\
 \bottomrule
 \end{tabular}
\end{table}

Frequency-mixing phenomena can be appreciated in \Figs{A-G_beam}(c) and \ref{A-G_beam}(d) as well. As discussed in Section \ref{F_mixing}, two different space-time modulations can lead to different output frequencies while maintaining the same diffraction angle. This is case for the diffraction orders ($m=1$, $n=1$) in \Fig{A-G_beam}(c) and ($m'=4$, $n'=4$) in  \Fig{A-G_beam}(d). Both diffraction orders have associated the same diffraction angle ($\theta_{11} = \theta_{4'4'} = 41.77^\mathrm{o} $) but different output frequencies ($\omega_{n=1} = 1.25 \omega_0$, $\omega_{n'=4} = 5 \omega_0$). In a general case,  eqs. \eqref{mixing1} and \eqref{mixing2} can be applied in the design of frequency mixers under oblique and normal incidence, respectively, by simply tuning the space-time parameters of the modulation. The key parameter in design is the time period, as the spatial period cannot be easily reconfigured electronically in the current metastructure. Without loss of generality, with this scenario, achieving the demodulation of the same signal by users located in different positions would be possible, which is highly interesting for applications in mobile or satellite communications. The present discussion opens the way to the design of spatiotemporal frequency mixers based on aperiodic metallic structures, which are expected to provide more degrees of freedom in design and ease of control than purely-periodic metallic ones \cite{taravati2018aperiodic}.


\subsection{Reflective Beamformer}
\begin{figure}[!t]
	\centering
    \includegraphics[width= 1\columnwidth]{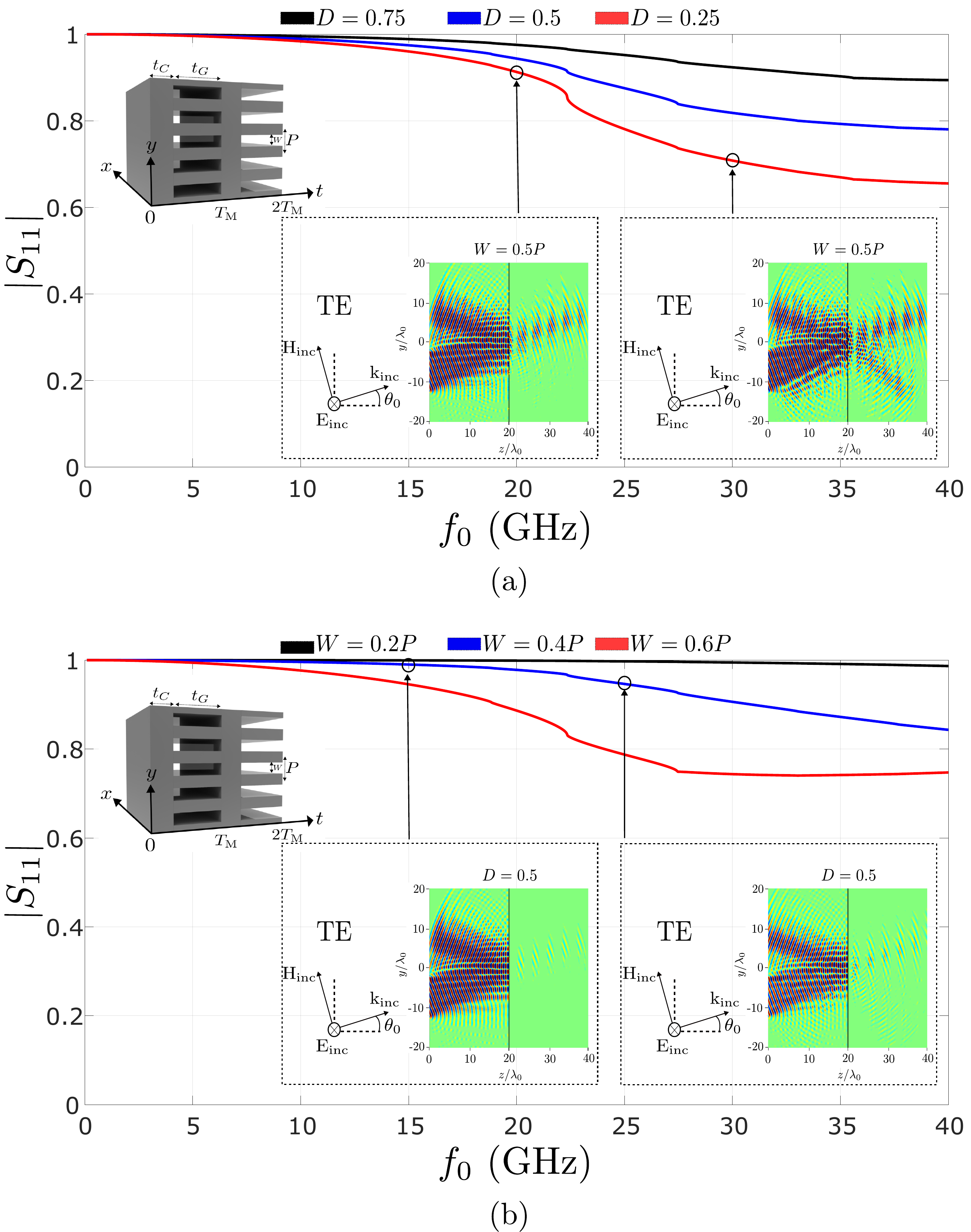}
	\caption{Reflection  coefficient as a function of the frequency of the incident wave for the C-G configuration. (a) When varying the duty cycle. (b) When varying the spatial dimension ratio.  TE oblique incidence $(\theta_0=20^\mathrm{o})$ is considered. Fixed parameters: $P=10$ mm, $T_{\text{M}}=4T_0$.}
	\label{C-G_beam}
\end{figure}

 In this subsection we test the tunability of the reflection coefficient for the C-G beamformer. \Fig{C-G_beam} depicts the analysis of the reflection coefficient as a function of the frequency of the incident wave when the temporal $T_{\text{M}}=4T_0$ ($f_0 = 4f_\text{M}$) and spatial $P=10$ mm periods are fixed. 
 Note that, in the present scenario, frequency of the modulation increases as the frequency of the incident wave does. Thus, the ratio $f_0 / f_\text{M}$ is fixed to a value of four.

The main advantage that the analytical approach contributes is the faster extraction of the scattering parameter for different points in frequency. Note that, the simulations carried out by means of the FDTD method [assuming a monochromatic impinging wave] takes several minutes depending on the mesh and the simulation space. However, the circuital approach takes only a few seconds to simulate the whole frequency range. 

 \Fig{C-G_beam}(a) shows the curves obtained by the proposed analytical method for the reflection coefficient $|S_{11}|$ when varying the duty cycle $D$. In this case, $DT_\text{M}$ represents the time in which the C-G metasurface is in the conductor state. FDTD simulations of the electric-field distributions are also included as insets. The black, blue and red solid lines are associated to the duty cycle  $D=0.75$, $D=0.5$ and $D=0.25$, respectively. It can be seen how the amplitude of the reflection coefficient decreases as the space-time metasurface remains less time in the conductor (fully reflective) state; namely, as $D$ decreases. This is in line with the former theoretical discussion and with the FDTD results inserted in \Fig{C-G_beam}(a).  Moreover, the oblique incident angle ($\theta_0=20^\mathrm{o}$) leads to the emerging of the grating lobe approximately at 23 GHz. This phenomenology is well caught by the circuital approach. It can be also appreciated in the FDTD results, as the grating lobes make appearance at 30 GHz (see the lower half space) but do not show at 20 GHz, below the grating-lobe regime. Finally, \Fig{C-G_beam}(b) shows the amplitude of the reflection coefficient in the case of varying the spatial ratio $W/P$ and fixing  the duty cycle. In this case, less reflection is observed as the spatial ratio (slit width $W$) increases. 
 
A crucial difference related to these two parameters should be noted. The modification of $D$ and $W$ is useful to adjust the reflection coefficient of the system, as explained in this section. However, as discussed in Section \ref{Harmonics}, the harmonic distribution of the diffraction pattern is also affected.  Nevertheless, the majority of $(m,n)$-th harmonics that can attain significant amplitudes when varying $W$ are of evanescent nature. This is not an issue from the beamforming perspective, since this parameter remains fixed in the grating state. On the other hand, when $D$ is modified, many of these $(m,n)$-th harmonics that can attain significant amplitudes will be propagative. Thus, this property of the temporal parameters turns them into the best option for adjusting the beamsteering of the system. Therefore, the best choice for setting the reflection and transmission coefficients of the system is selecting the proper configuration (A-G or C-G), as expected. This feature is very interesting for future wireless communications where the communication at both sides of the device is desired \cite{zhang2021intelligent}.

Finally,  the efficiency of the proposed analytical method is noteworthy. The computational times associated with our analytical (circuit-based) approach are much lower compared to other numerical techniques. Computation time is less than a second here, while FDTD and other numerical approaches normally take minutes to provide the solution. From a practical perspective, our approach is an efficient method that can be used as a surrogate model to be combined with external optimization or artificial intelligence techniques for an efficient design of future metal-based space-time metastructures.

Although the experimental part is beyond the scope of the paper, it is worth mentioning some of the challenges that a future implementation of the space-time metasurface will face. Naturally, an interesting alternative in the microwaves is the use of PIN and varactor diodes.  Some alternatives of this kind can be found in the literature. Switching times and related transients between ON and OFF states in PIN diodes and varactors are indeed one of the main technological challenges. From the theoretical side, we are extending the analytical framework to include the complete circuit model of these active elements \cite{perez2023analytical} (including parasitic effects at higher frequencies).  In the low-THz regime, a more exotic alternative could be the use of graphene. Blue and red strips in Fig. 1 could be treated as biased elements at different voltages, thus giving the air and conductor responses. Optical pumping could be an alternative too \cite{tasolamprou2019experimental,he2021ultrafast}. The performance of a future prototype will then be validated experimentally in an anechoic chamber and an optic table. The main task will be to check the beamsteering and frequency mixing capabilities of the space-time metasurface by measuring the transmission and reflection coefficients and the diffracted angles with any of the combinations of the three states: air, conductor and grating.

\section{Conclusions}
In this manuscript, we have introduced a novel metal-based spatiotemporal metasurface for application in wireless communications systems. The metasurface can alternate its properties in time within three different states: "air" (fully transparent), "conductor" (fully reflective) and "grating" (partially transparent and reflective). By combining the three states in a time-periodic scheme, a rich diffraction pattern is created, which can be electronically reconfigured by setting the space-time parameters of the system.

The physics of the space-time metasurface is described by means of an analytical technique based on equivalent circuits, Floquet-Bloch expansions and integral equations. We have considered  scenarios where the metasurface is illuminated by oblique TE and TM plane waves. The analytical technique has proven to be computationally efficient compared to general numerical approaches and the self-implemented FDTD code. Moreover, it provides physical insight on the diffraction spectrum, nature (evanescent/propagative) of the space-time Floquet harmonics, and scattering parameters. 

The results of the work show that the present space-time metasurface offer clear advantages compared to our previous time-only configuration. Here, we have exploited
the inherent ability of the space-time metasurface to mix and multiply frequencies. We have shown that two or more output frequencies can be engineered so their diffraction angles is the same. Finally, we have shown that efficient beamforming can be realized by tuning the space-time parameters of the system. Combinations of the conductor and grating states are prone to smartly reflect most of the diffracted power. Conversely, combinations of the air and grating states can enable effiecient beamforming in transmission. 

\section*{Acknowledgments}
This work has been supported by grant TED2021-129938B-I00 funded by MCIN/AEI/10.13039/501100011033 and by the European Union NextGenerationEU/PRTR. It has also been supported by grants PID2020-112545RB-C54, PDC2022-133900-I00, and PDC2023-145862-I00, funded by MCIN/AEI/10.13039/501100011033 and by the European Union NextGenerationEU/PRTR. It is also part of the grant IJC2020-043599-I funded by MICIU/AEI/10.13039/501100011033 and by European Union NextGenerationEU/PRTR.. 

\section*{Appendix A}
The analytical results have been compared with a self-implemented 2D finite-difference time-domain (FDTD) method based on a staggered grid scheme.  Two independent codes have been programmed for TE and TM incidences. For an impinging TE-polarized wave ($E_x, H_y, H_z$), the space-time-varying boundary conditions in the metallic regions of the metasurface (placed in $z_{pos}$)  are enforced as $E_x=0$  in both conductor and grating states. For TM incidence ($ E_y, E_z, H_x$), the space-time-varying boundary conditions in the metallic regions are enforced as $E_y=0$. Second-order Engquist-Majda relations have been implemented as absorbing boundary conditions at the outer limits of the simulation box. The rest of parameters that are involved in the FDTD simulations are: uniform mesh grid with $\Delta y=\Delta z =\frac{\lambda_0}{35}$; time step of $\Delta t=C \cdot \frac{\Delta z}{c}$, with $C=0.5$ being the CFL stability condition parameter. To model the zero-thickness metasurface, a homogeneous slab with a thickness less than $0.1  \lambda_0$ has been considered.

\bibliography{references}

\end{document}